\documentclass[aip,
amsmath,
amssymb,
graphicx,
pop,
reprint,
]{revtex4-1}

\usepackage{graphicx}
\usepackage[utf8]{inputenc}
\usepackage[T1]{fontenc}

\newcommand{\m}{\text{m}}

\newcommand{\mev}{\text{MeV}}
\newcommand{\gev}{\text{GeV}}

\newcommand{\cm}{\text{cm}}

\newcommand{\mum}{\text{$\mu$}\m}

\newcommand{\joule}{\text{J}}

\newcommand{\fs}{\text{fs}}

\newcommand{\fwhm}{\text{FWHM}}

\begin{document}

\preprint{AIP/123-QED}
\title{Limitations of post accelerating ion beams using the snowplow field in a near-critical density target} 



\author{Davide Terzani}
\email[]{dterzani@lbl.gov}
\affiliation{Lawrence Berkeley National Laboratory, Berkeley, CA 94720, USA}
\author{Stepan S. Bulanov}
\affiliation{Lawrence Berkeley National Laboratory, Berkeley, CA 94720, USA}
\author{Lieselotte Obst-Huebl}
\affiliation{Lawrence Berkeley National Laboratory, Berkeley, CA 94720, USA}
\author{Carlo Benedetti}
\affiliation{Lawrence Berkeley National Laboratory, Berkeley, CA 94720, USA}
\author{Franklin Dollar}
\affiliation{University of California, Irvine, CA}
\author{Eric Esarey}
\affiliation{Lawrence Berkeley National Laboratory, Berkeley, CA 94720, USA}
\author{Axel Huebl}
\affiliation{Lawrence Berkeley National Laboratory, Berkeley, CA 94720, USA}
\author{Aodhan McIlvenny}
\affiliation{Lawrence Berkeley National Laboratory, Berkeley, CA 94720, USA}
\author{John Palastro}
\affiliation{University of Rochester, Laboratory for Laser Energetics, Rochester, NY}
\author{Jessica Shaw}
\affiliation{University of Rochester, Laboratory for Laser Energetics, Rochester, NY}
\author{Carl B. Schroeder}
\affiliation{Lawrence Berkeley National Laboratory, Berkeley, CA 94720, USA}
\affiliation{Department of Nuclear Engineering, University of California, Berkeley, CA 94720, USA}
\author{Douglass Schumacher}
\affiliation{Ohio State University, Columbus, OH}
\author{Mingsheng Wei}
\affiliation{University of Rochester, Laboratory for Laser Energetics, Rochester, NY}
\author{Louise Willingale}
\affiliation{The Gérard Mourou Center for Ultrafast Optical Science, University of Michigan, Ann Arbor, MI 48109}

\date{\today}

\begin{abstract}
Laser-matter interaction at relativistic intensities is central to almost every scientific case for multi-PW laser facilities.
In particular, the progress in laser-driven ion acceleration brings this interaction closer to multiple applications,
ranging from material science to biomedical research. One of the most important questions regarding ion acceleration is how to obtain high charge ion beams with relativistic energies.
Increasing laser energy and intensity usually leads to a number of limitations on maximum achievable ion energies rooted in the fundamental properties of charged particle interaction with strong electromagnetic fields.
Following advances in staged laser-plasma electron acceleration, staged ion acceleration
could offer a path to relativistic energies.
Here, such scheme is explored and the possibility of scaling the acceleration of ion beams to relativistic energies is discussed.
\end{abstract}

\pacs{}

\maketitle 

\section{Introduction}

Laser-matter interaction at relativistic intensities has attracted considerable attention from the scientific community for more than half a century.
It started as a search for a basic understanding of the fundamental physics of plasma waves and
then progressed to the study of different plasma instabilities, particle acceleration, magnetic reconnection, and high-frequency radiation generation~\cite{mourou_optics_2006,esarey_physics_2009,daido_review_2012,macchi_ion_2013,piazza_multi-petawatt_2022}.
A number of applications of laser-matter interaction with significant societal impact include extreme-ultraviolet lithography
for producing integrated circuits for the semiconductor industry,
compact Laser-Plasma Accelerators (LPAs) for generating electron and ion beams and serving as light sources,
and the demonstration of fusion gain at the National Ignition Facility.
As the available laser intensity increases, new phenomena can be studied, as well as novel, groundbreaking applications,
including laser-plasma-based sources of relativistic ions.
In what follows, we study a scheme for laser-ion acceleration that depends on laser-matter interaction at relativistic intensities. 

Recently, there has been a lot of interest in studying high-intensity, high-power laser interactions with near-critical-density (NCD) targets,
defined as targets whose density is comparable to, or slightly greater than, the critical plasma density
(see, e.g., Refs.~\onlinecite{zhu_dense_2016,stark_enhanced_2016,sharma_high_2018,psikal_laser-driven_2021,he_achieving_2022,garten_laser-plasma_2024,hakimi_lasersolid_2022,rehwald_ultra-short_2023,seemann_laser_2024}).
The critical plasma density is defined as the threshold between the plasma being opaque and being transparent to laser light,
and can be expressed as $n_{cr}=m_e \omega_0^2/4\pi e^2\simeq 1.12\times 10^{21} \cm^{-3}/\lambda_0^2\,[\mum],$ where $m_e$ is the electron mass, $e$ is the electron charge, $\omega_0=2\pi c/ \lambda_0$
is the laser frequency, and $\lambda_0$ is the laser wavelength, respectively. 
As the laser pulse intensity increases, the transparency threshold increases as well due to the relativistic
effects~\cite{daido_review_2012,macchi_ion_2013}, so that opaque targets become transparent for the laser radiation.
During the interaction with an NCD target, the laser pulse can be almost completely depleted, with a significant amount of laser energy absorbed by plasma electrons and ions.
The collective motion of electrons and ions leads to the generation of extreme electromagnetic fields in plasma,
whose strength might approach mega-Tesla and peta-eV/m according to different analytical
and computer simulation estimates~\cite{stark_enhanced_2016,park_ion_2019,rinderknecht_relativistically_2021,hakimi_lasersolid_2022,hakimi_dephasing_2024}.
Plasma density modulations excited during the interaction move with different velocities and in different directions,
fueling a number of instabilities (see, e.g., Refs.~\onlinecite{bulanov_ion_2012,haberberger_collisionless_2012}).
Laser pulse propagation in a NCD target has been studied for electron and ion acceleration
and for the generation of high-energy photons~\cite{bulanov_generation_2010,willingale_collimated_2006, willingale_high-power_2011,gong_energy-chirp_2022,stark_enhanced_2016,vranic_extremely_2018,babjak_direct_2024,gong_laser_2024}.

The intense laser pulse volumetric interaction with the NCD plasma in the regime of relativistic transparency
requires PW or multi-PW class laser facilities, such as ELI-NP~\cite{radier_10_2022, tanaka_current_2020},
ELI Beamlines~\cite{spinka_commissioning_2017, weber_p3_2017}, CoReLS~\cite{sung_42_2017},
Apollon~\cite{yao_characterization_2025}, SULF~\cite{li_339_2018}, ZEUS~\cite{maksimchuk_zeus_2025},
and the future NSF OPAL~\cite{bromage_technology_2019}.
Almost all these facilities either deliver or plan to deliver two or more synchronized PW and multi-PW laser pulses to the target chamber,
e.g., ELI-NP with two 10 PW pulses, or the future NSF OPAL facility with two 25 PW pulses.
Such multi-beam configurations enable exciting possibilities to study laser-matter interactions at relativistic intensities with NCD targets.
Among all of these, two pulses can be used to demonstrate a two-stage ion acceleration.
In this scheme, the ions generated by the first laser pulse are injected into the second stage, there interacting with a second NCD target.
Their energy is then boosted by the strong electromagnetic fields generated in front of the laser pulse at the laser-plasma interface,
known as the "snowplow" regime of laser-ion acceleration~\cite{shorokhov_ion_2004, liu_front_2020, liu_accelerating_2022}
(for similar approaches to multi-stage laser ion acceleration see Ref.~\onlinecite{garten_laser-plasma_2024} and references cited therein).

In this paper, we study the operation of a second stage of a two-stage laser ion accelerator,
where both stages are powered by multi-PW laser pulses.
To assess the underlying physics of ion acceleration by the electromagnetic fields generated in a multi-PW laser interaction with an NCD target,
we adopt a simplified probing model.
An externally injected test proton beam with a variable time delay relative to the second pulse is used.
The beam has a 100\% energy spread and a maximum energy of $2\,\gev$, which provides an upper bound estimate for the magnetic vortex acceleration regime~\cite{park_ion_2019}.
This choice enables us to examine its initial phase space in full.
The beam propagates collinearly with the laser and is assumed to have zero angular divergence.
This initial condition is chosen to simplify the modeling and enable a comparison with 1D analytical theory.
We compute the resulting energy boost as a function of the initial proton phase-space coordinates
to infer the accelerating fields inside the NCD target and identify conditions that may optimize the two-stage acceleration.
This also allows us to examine the maximum energy gain,
which is governed by the laser-plasma interface (snowplow) velocity~\cite{shorokhov_ion_2004, liu_front_2020, liu_accelerating_2022}.
The simplified model is validated using Particle-In-Cell (PIC) simulations.

This paper is organized as follows: in Section~\ref{sec:laserpropagation} we estimate externally injected proton beam energy gain due to the snowplow field.
These estimates are benchmarked against the results of PIC simulations in Section~\ref{sec:simulations}.
We discuss the obtained insights into the post-acceleration of proton beams using snowplow field in an NCD target in Section~\ref{sec:conclusions}. 

\section{Laser pulse propagation in near critical density plasma}\label{sec:laserpropagation}

In this Section, we use an analytical model to estimate the energy gain of
pre-accelerated protons in the snowplow field of the laser pulse propagating in an NCD plasma.
The accelerating electric field is generated at the laser-plasma interface and moves with it.
Thus, in order to estimate the energy gain we need to analyze the laser propagation in the NCD plasma
and how the laser-plasma interface velocity is connected to the laser and plasma properties.
This propagation can be described in the framework of the waveguide model~\cite{bulanov_helium-3_2015,park_ion_2019},
where it is assumed that an intense laser pulse impinging on a NCD target produces a density channel,
by first displacing the electrons, followed by the ions.
The channel walls have density high enough to contain the laser propagation and reduce its diffraction.
The laser propagates inside this self-generated channel, so that its electromagnetic field can be described by a wave equation in a waveguide.
It is possible to use the solution of the wave equation in a waveguide~\cite{hakimi_dephasing_2024} to determine
the the radius of the channel, $R_{ch}$, as well as the laser field strength, $a_{ch}=eA_{ch}/m_ec^2$, where $A_{ch}$
is the amplitude of the vector potential of the electromagnetic field in the channel
and the group velocity, $\beta_g=\sqrt{1-1/\gamma_g^2}$ as:
\begin{align}
&R_{ch}=\frac{\lambda_0}{\pi}\tilde{n}_e^{-1/3}\tilde{P}^{1/6}, \label{eq:waveguide_radius}\\
&a_{ch}=\tilde{n}_e^{1/3}\tilde{P}^{1/6}, \label{eq:waveguide_a}\\
&\gamma_g=\frac{\sqrt{2}}{1.84}\tilde{P}^{1/6}\tilde{n}_e^{-1/3}.\label{eq:waveguide_gamma}
\end{align}
In Equations~\eqref{eq:waveguide_radius}-\eqref{eq:waveguide_gamma},
we introduced the normalized laser power, $\tilde{P}=(2/K)(P/P_c)$,
and normalized plasma density, $\tilde{n}_e=n_e/n_{cr}$.
Here $P_c=2m_e^2c^5/e^2=17$ GW is the characteristic power for relativistic self-focusing, $K=(\pi/32)^{1/2}[J_1(\kappa R_{ch})^2-J_0(\kappa R_{ch})J_2(\kappa R_{ch})]\simeq1/13.5$
is a geometrical factor coming from the integration over the laser pulse profile in the channel~\cite{bulanov_generation_2010},
$J_n(\kappa R_{ch})$ is the Bessel function of the first kind, and $\kappa\simeq 1.84/R_{ch}$ is the solution of the equation
$J^\prime_1(\kappa R_{ch})=0$, which is a boundary condition for the wave equation in the waveguide.

\begin{figure}[!ht]
\centering
\includegraphics[width=8.6cm]{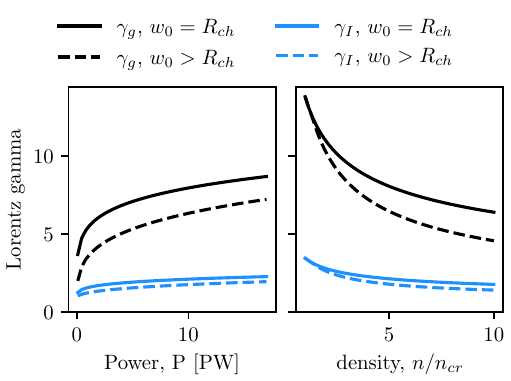}
\caption{The dependence of the laser $\gamma_g$ (black curves) and $\gamma_I$ (blue curves) as a function of: (left panel) laser power, with fixed plasma density, $n_e/n_{cr}=4$, (right panel) plasma density, with fixed laser power, $P=17$ PW. The case with the effect of $w_0>R_{ch}$ is shown by dashed curves. Here, $w_0=4\lambda_0$.}
\label{fig:interface gamma}
\end{figure}

As the laser pulse propagates in the NCD plasma, it loses energy due to the generation of the channel in both electron and ion plasma components,
which is accompanied by significant energy gain by plasma particles.
This energy depletion is characterized by a depletion length, $L_{ch}$, which can be determined as follows.
As the energy of the laser in the channel, $W_L=\pi R_{ch}^2 T_{\fwhm} a^2_{ch}m_e c^3 n_{cr}K$, where $T_{\fwhm}$ is the full width at half maximum laser duration, is spent mostly on electron acceleration,
then by assuming that the average energy gain by electrons initially placed in this channel of radius $R_{ch}$ and length $L_{ch}$ is $a_{ch}m_e c^2$
and equating the laser energy to the energy gained by these electrons,
$W_e=\pi R_{ch}^2 L_{ch} n_e a_{ch} m_e c^2$, we obtain the depletion length, which is also the channel length:
\begin{equation}
\frac{L_{ch}}{L_p}=A\gamma_g^2,
\end{equation}
where $A \simeq 1.84^2K/2 \simeq 0.125$ and $L_p=cT_{\fwhm}/2\sqrt{\log(2})$ is the laser pulse length. If we assume that the laser pulse depletes homogeneously, i.e., the rate of energy transfer from the laser to the plasma electrons is constant, the velocity of the laser-plasma interface is 
\begin{equation}
\beta_I=\frac{L_{ch}}{L_{ch}+L_p}\beta_g.
\label{eq:beta_interface}
\end{equation}
The Lorentz factor associated with the motion of the laser-plasma interface is
\begin{equation}
    \gamma_I=\frac{1+A\gamma_g^2}{\sqrt{1+\gamma_g^2(A^2+2A)}}.
    \label{eq:gamma_interface}
\end{equation}
Thus, the Lorentz factor of the laser-plasma interface is determined by the group velocity of the laser.
According to Eq.~\eqref{eq:gamma_interface}, it varies from unity, when $\gamma_g=1$, to $\sim \gamma_g/4$, when $\gamma_g\gg 1$.
For values of plasma density and laser power that satisfy the validity requirements
of the waveguide model~\cite{bulanov_generation_2010}, this variation is even smaller (see Fig.~\ref{fig:interface gamma}).
Moreover, the weak dependence of $\gamma_g$ on $\tilde{P}$ and $\tilde{n}_e$ shown in Eq.~\eqref{eq:waveguide_gamma},
leads to a Lorentz factor of the laser-plasma interface varying by a factor of two over the order of magnitude
variation of either laser power (Fig.~\ref{fig:interface gamma}, left panel) or plasma density (Fig.~\ref{fig:interface gamma}, right panel).
Thus, the weak dependence of the $\gamma_g$ on laser power and plasma density as well as a significant reduction
of the Lorentz factor of the laser-plasma interface compared to $\gamma_g$,
which is due to strong depletion of laser pulses in the NCD plasma,
are the main limiting factors for the energy gain in the two stage ion acceleration scheme based on the snowplow regime.
In what follows we show that a several GeV energy gain is potentially possible in the framework of this acceleration scheme,
however scaling the scheme up 10 GeV and beyond is prohibitively challenging and would require a different approach to boosting the energy of relativistic ions.

The results above were obtained under the assumption of an ideal coupling, i.e.,
laser pulse width at focus being equal to the radius of the self-generated channel, $w_0=R_{ch}$~\cite{park_ion_2019,hakimi_dephasing_2024}.
However, in any realistic experimental scenario the laser pulse focal radius will be fixed,
whereas the density of the target and laser pulse power can be varied, resulting in $w_0\neq R_{ch}$ in general.
For the case $w_0>R_{ch}$, when only a fraction of the laser power is effectively coupled into the self-generated channel,
different expressions for $R_{ch}$, $\gamma_g$, and $L_{ch}$ should be used~\cite{hakimi_dephasing_2024}:
\begin{align}
    &R_{ch}=\frac{\lambda_0}{\pi}\left(\frac{\lambda_0}{\pi w_0}\right)^{1/2}\left(\tilde{n}_e^{-2}\tilde{P}\right)^{1/4}, \label{eq:rch_coupling} \\
    &L_{ch}/L_p=K\frac{\lambda_0}{\pi w_0}\left(\tilde{n}_e^{-2}\tilde{P}\right)^{1/2}, \label{eq:Lch_coupling}\\
    &\gamma_g=\frac{\sqrt{2}}{1.84}\left(\frac{\lambda_0}{\pi w_0}\right)^{1/2}\left(\tilde{n}_e^{-2}\tilde{P}\right)^{1/4}, \label{eq:gamma_group_coupling}
\end{align}
which result in a different behavior for $\gamma_I$,
when the plasma density is varied (see Fig.~\ref{fig:interface gamma}).
The part of the laser that is outside the $\pi R_{ch}^2$ area of the channel
is scattered in plasma and does not participate in creating the channel and in driving the snowplow field.
Here it is assumed that the laser power is homogeneously distributed across the focal spot and only
$P_{ch}=(R_{ch}/w_0)^2P$ part of initial laser power is coupled to the channel.
As a result, the non-ideal case demonstrates stronger dependence on plasma density leading to smaller values of laser-plasma interface velocity.

In what follows, we use the results above to estimate the energy gain of the proton beams
accelerated by the longitudinal electric field associated with the laser-plasma interface.
\begin{figure}[!ht]
\centering
\includegraphics[width=8.6cm]{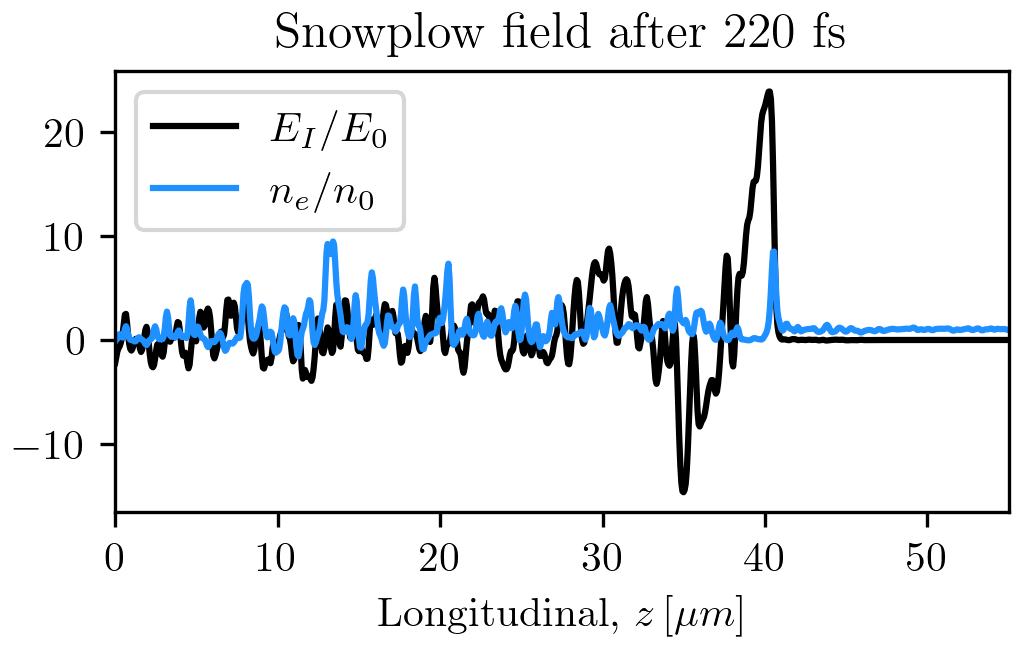}
\caption{On-axis lineout of the snowplow electric field and density
at $t=220\,\fs$ generated by a laser with energy $\mathcal{E}=500\,\joule$ in a density density $n_0=2n_{cr}$.
The other laser and plasma parameters are listed in table~\ref{tab:simulation_parameters}.
For this density, the cold wavebreaking limit is $E_0\simeq 5.7\,\text{TV/m}.$}
\label{fig:snowplow_longitudinal_field}
\end{figure}
Figure~\ref{fig:snowplow_longitudinal_field} shows the longitudinal charge-separation field (black line)
and background electron density (blue line) for a $\mathcal{E}=500\,\joule$ laser pulse propagating
in a $\tilde{n}_e=2$ target after a $220\,\fs$ propagation.
The laser front (omitted for clarity) sits at $z\gtrsim 40\,\mum$, where it pushes the plasma density.
Particles ahead of the moving field interface are accelerated by the interface itself until
their momentum becomes large enough to outrun it.
For brevity, we do not derive here the full analysis of particle motion in the interface field,
which can be found in several works, for instance in Refs.~\onlinecite{shorokhov_ion_2004, brantov_synchronized_2016, liu_accelerating_2022, gothel_optimized_2022}.
The equations of motion for these particles can be written in the following form:
\begin{align}
    &\frac{du}{dt}=\frac{m_e}{m_p}E_I(x-\beta_I t),\label{eq:eq_motion_x_snowplow}\\
    &\frac{1}{c}\frac{dz}{dt}=\frac{u}{(1+u^2)^{1/2}}.\label{eq:eq_motion_p_snowplow}
\end{align}
Here, $z$ is the longitudinal coordinate, $u$ is the relativistic momentum normalized to $m_pc$, with $m_p$ the proton mass, 
and $E_I(x-\beta_I t)$ is the longitudinal charge-separation field, moving with the interface and normalized to $E_0=m_ec\omega_p/e$,
where $\omega_p=\sqrt{4\pi e^2 n_e/m_e}$ is plasma frequency.
In the reference frame co-moving with the interface, the particle acceleration in the interface field can be interpreted as a reflection
from a relativistic mirror \cite{bulanov_relativistic_2013}.
Particles that move faster than the interface do not interact with the moving structure.
Conversely, particles that move slower than the interface are swept up by the field.
Because the field has a finite longitudinal extent (as shown in Figure~\ref{fig:snowplow_longitudinal_field}),
they experience net acceleration over that extent.
If the initial particle momentum is high enough,
the energy acquired during the interaction is sufficient to trap the particles in the structure rather than allowing them to slip back.
In this case, the relativistic mirror reflects them forward in its reference frame.
The closer the initial speed is to the lower trapping threshold, the longer the particles interact with the field,
and the higher their final velocity.

The Hamiltonian analysis in Ref.~\onlinecite{liu_accelerating_2022} provides an analytical solution of the equations under a 1D approximation.
For higher target densities, the interface velocity is smaller, which lowers the initial trapping threshold.
However, the maximum energy gain is limited by the shorter interaction time associated with a smaller field extent and the smaller velocity of the interface field.
For lower target densities, the trapping threshold increases because the interface moves faster and particles can slip through the field more easily.
Nevertheless, the faster-moving interface covers a larger physical extent, which can imprint higher energy
gains on the particles that remain trapped and thus reach higher final energies.

\section{Simulations}\label{sec:simulations}

We investigated the action of the snowplow field on pre-accelerated test protons
using a set of PIC simulations performed with WarpX~\cite{fedeli_pushing_2022, vay_warp-x_2018}.
The laser-plasma interaction was modeled in quasi-cylindrical symmetry, in which the electromagnetic fields are decomposed into azimuthal modes.
Compared with fully 3D simulations, this approach substantially reduces the computational cost while preserving the correct scaling of the electromagnetic fields in the target,
which is not generally captured in a 2D Cartesian geometry.

The simulations used a fixed window that included the target in its entirety.
The longitudinal and transverse resolutions were 64 and 51 cells per micron, respectively.
The background electron-proton plasma was represented by 70 macroparticles per cell for each species.
The input files used for the simulation scan can be found in Ref.~\onlinecite{terzani_limitations_2026}.
\begin{table}[h!]
\begin{tabular}{r|l}
Parameter & Value \\
\hline
\hline
\multicolumn{1}{l}{\textbf{Laser-plasma}} & \\
Energy, $\mathcal{E}\,[J]$ & 200, 500 \\
Strength, $a_0$ & 76, 120 \\
Duration, $T_{\fwhm}\,[\fs]$ & 30 \\
Waist at Focus, $w_0,\,[\mum]$ & 4 \\
Wavelength, $\lambda_0,\,[\mum]$ & 0.8 \\
Initial position, $z_0\,[\mum]$ & -28 \\
Target Density, $n_e/n_c$ & 1 - 10 \\
\multicolumn{1}{l}{\textbf{Test protons}} & \\
Energy, $\mathcal{E}_i\,[\mev]$ & 0.75 - 2030 \\
Transverse Position, $r\,[\mum]$ & -0.5 - 0.5 \\
Longitudinal Position, $z\,[\mum]$ & -30 - 30 \\
\end{tabular}
\caption{Simulation Parameters}
\label{tab:simulation_parameters}
\end{table}

The simulations model a pre-ionized, near-critical, pure hydrogen target.
To isolate the snowplow acceleration mechanism, we introduce pre-accelerated protons as test particles,
thereby representing a proton population produced by an unspecified preceding acceleration stage.
The details of this pre-acceleration stage, as well as the mechanism by which protons couple to the snowplow field, are beyond the scope of this study.

The test particles are initialized over a broad region of longitudinal phase space, allowing us to determine the final proton energy as
a function of their initial position and energy.
Their initial energy range from $\mathcal{E}_i=0.75\,\mev$ to $\mathcal{E}_i=2030\,\mev$,
while their longitudinal positions span $z=-30\,\mum$ to $z=30\,\mum$.
All protons are initially injected with zero transverse momentum and are distributed transversely over
$r=-0.5\,\mum$ to $r=0.5\,\mum$ around the laser axis.

Restricting the initial transverse extent to this narrow region, which is substantially smaller than both the laser waist and the plasma channel formed in the target,
simplifies the analysis while retaining the essential features of the acceleration process.
In particular, the initially collinear distribution reduces transverse--longitudinal coupling and transverse phase-space mixing,
enabling a more direct comparison with the 1D analytical model.

We considered a circularly polarized laser pulse with either $200\,\joule$ or $500\,\joule$ of energy.
The pulse is characterized by a Gaussian longitudinal and transverse profile.
The other laser parameters are described in Table~\ref{tab:simulation_parameters}.
The normalized laser strengths for the two pulses are $a_0 = 76$ and $a_0 = 120$ respectively.
Our simulations explore a range of target densities, normalized to the critical density $n_c$, from 1 to 10 in integer steps.
By varying the target density and laser energy, we investigate the effects of these parameters on the snowplow acceleration process.
The collection of parameters for each simulation are summarized in Table~\ref{tab:simulation_parameters}.
\begin{figure}[!ht]
\centering
\includegraphics[width=8.6cm]{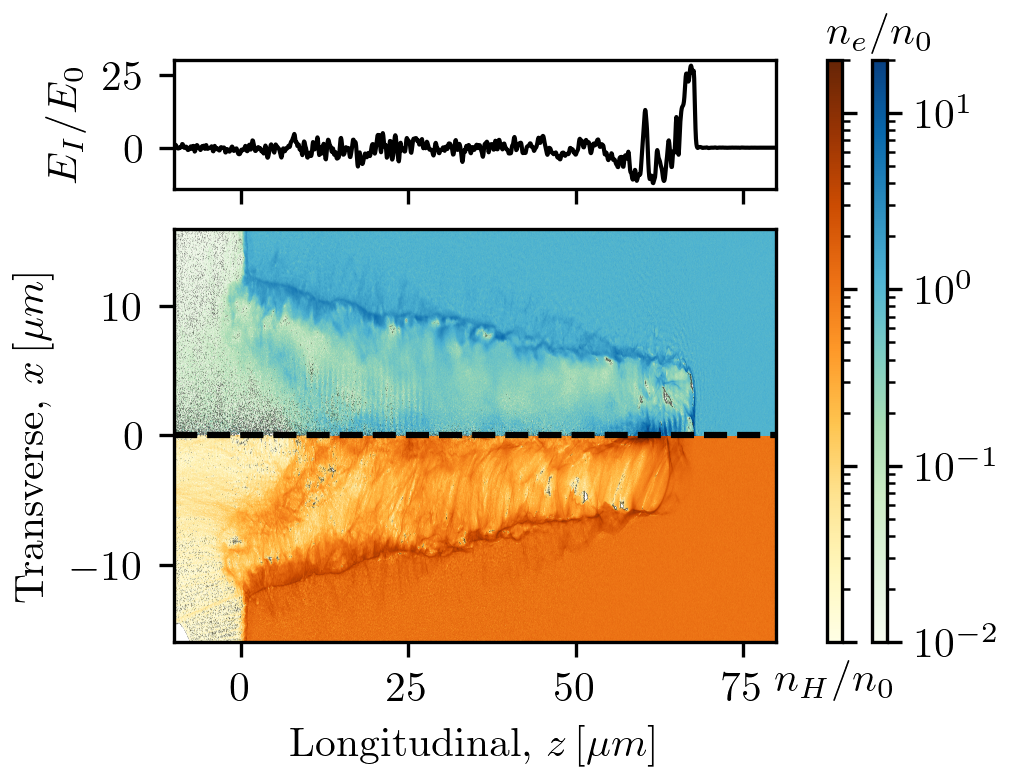}
\caption{The bottom panel shows the electron (top, blue scale) and hydrogen (bottom, red scale)
density maps after $t=320\,\fs$ propagation.
The top panel shows the corresponding normalized longitudinal electric field on axis.
The dynamics show the interface at about $70\,\mum$, where the laser front is pushing the electrons.
The ions react with some delay due to their larger mass.
The incoming laser energy is $\mathcal{E}=500\,\joule$ and the unperturbed density is $\tilde{n}_e=2$.
The other laser and plasma parameters are listed in table~\ref{tab:simulation_parameters}.
For this density, the cold wavebreaking limit is $E_0\simeq 5.7\,\text{TV/m}.$}
\label{fig:ele_hyd_density_map}
\end{figure}

As mentioned above, the propagation of the intense laser pulse in the NCD plasma leads
to the formation of laser generated channels in electron and ion plasma components as shown in Figure~\ref{fig:ele_hyd_density_map},
where we show a density map of the electrons (top half) and ions (bottom half) at $t = 320\,\fs$ for a simulation with a 500 J laser pulse and a target density $\tilde{n}_e=2$.
The laser pulse propagates from left to right.
The formation of the channel is clearly seen, as well as the fact that the channel in ion density is formed
with a slight delay after the the channel in electron density.
\begin{figure}[!ht]
\centering
\includegraphics[width=8.6cm]{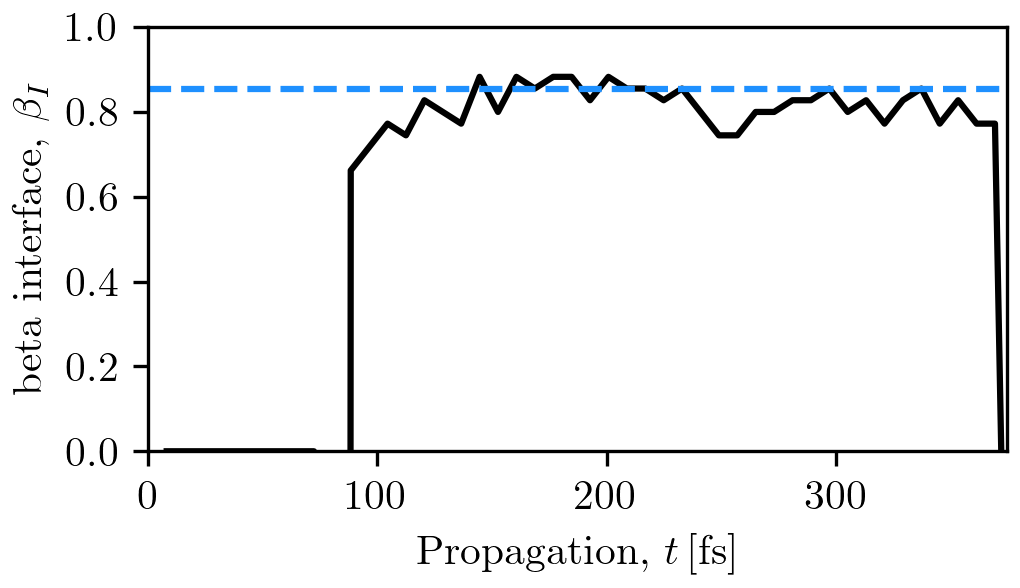}
\caption{The evolution of the laser-plasma interface velocity,
$\beta_I$, for a  $\mathcal{E}=500\,\joule$ laser pulse propagating in a $\tilde{n}_e=4$ density plasma.
The dashed line shows the value predicted by Eq.~\eqref{eq:beta_interface}
for these laser and plasma parameters.
}
\label{fig:beta_int}
\end{figure}

The evolution of the laser-plasma interface velocity, $\beta_I$, for a
$\mathcal{E}=500\,\joule$ laser pulse propagating in the $\tilde{n}_e=4$ plasma is shown in Figure~\ref{fig:beta_int}.
At about $t\simeq 80\,\fs$, when a substantial fraction of the laser pulse enters into the target,
$\beta_I$ reaches approximately 0.8.
The subsequent decrease around $t\simeq 400\,\fs$ indicates almost total laser pulse depletion.
During its propagation, the laser pulse travels about $80\,\mum$ into the target.
The analytical estimate for the interface velocity given by Eq.~\eqref{eq:beta_interface}
agrees reasonably well with the maximum value of the velocity.
\begin{figure}[!ht]
\centering
\includegraphics[width=8.6cm]{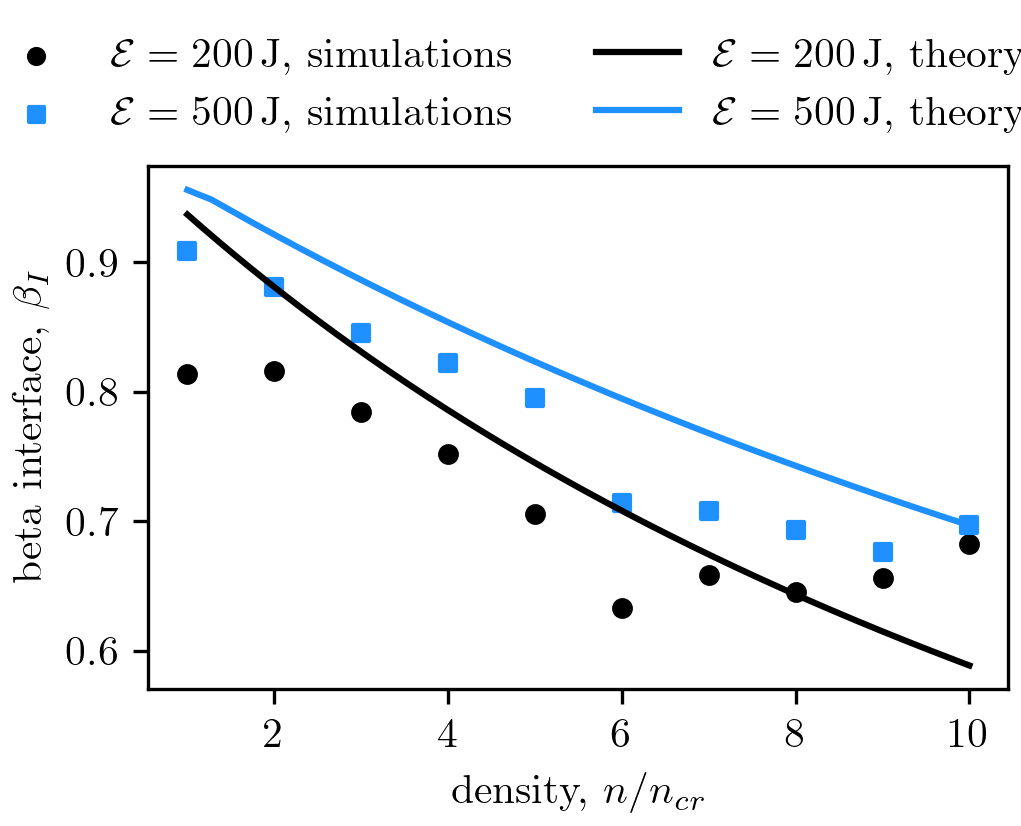}
\caption{The dependence of the average value of laser-plasma interface velocity on plasma density for two values of laser energy,
$\mathcal{E} = 200\,\joule$ (black curve and dots) and $500\,\joule$ (blue curve and dots).
The curves are the analytical results from Eq.~\eqref{eq:beta_interface}. Dots are the results of PIC simulations.}
\label{fig:beta_average}
\end{figure}

The laser plasma interface velocity depends on the laser power and on plasma density.
With the increase of the latter, the laser plasma interface velocity decreases,
since the laser depletes faster in more dense plasma,
which can be seen in Figure~\ref{fig:beta_average}, where the dependence of $\beta_I$
averaged over the laser depletion length is shown as a function of plasma density for two values the laser energy.
The solid lines represent the scaling expected according to the simplified model presented in Eqs.~\eqref{eq:beta_interface}~-~\eqref{eq:gamma_group_coupling}.
The results of PIC simulations are in reasonable agreement with the analytical estimates.
\begin{figure}[!ht]
\centering
\includegraphics[width=8.6cm]{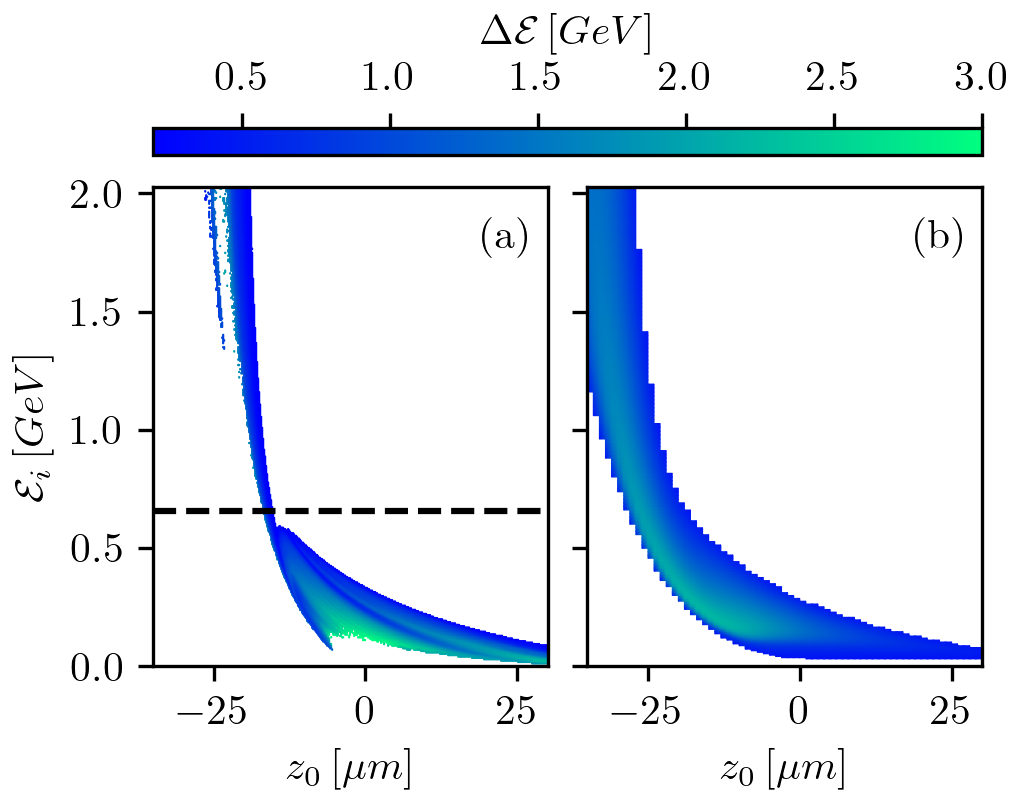}
\caption{The dependence of the proton energy gain, $\Delta \mathcal{E}$
on the initial longitudinal position $z_0$ and initial kinetic energy $\mathcal{E}_i$
for a 500 J laser pulse and a target density of 4$n_c$.
The other laser and plasma parameters are listed in table~\ref{tab:simulation_parameters}.
The left panel (a) shows the results obtained via PIC simulations,
the right panel (b) shows the numerical solutions of equations~\eqref{eq:eq_motion_x_snowplow}-\eqref{eq:eq_motion_p_snowplow}.
The maximum energy gain is achieved for particles starting ahead of the laser-plasma interface with energies $200\,\mev\lesssim \mathcal{E}_I \lesssim 500\,\mev$
and swept by the snowplow field.
The horizontal dashed line corresponds to the energy of a proton initially moving with the velocity equal to the average interface velocity.
The plots mask phase space regions corresponding to particles losing energy in the interaction.}
\label{fig:energy_gain}
\end{figure}

The test protons are being accelerated by the snowplow field that is generated at the laser-plasma interface,
i.e., in front of the laser pulse.
However, the total energy gain of these protons depends on the initial conditions, i.e., on their initial position and momentum.
In order to characterize this acceleration we present a color map of the energy gain
$\Delta \mathcal{E}$ of individual test protons as a function of their initial longitudinal position $z_0$
and initial kinetic energy $\mathcal{E}_i$ for a simulation with a 500 J laser pulse and a target density of $4\,n_c$ (see the left panel of Figure~\ref{fig:energy_gain}).
The horizontal dashed line corresponds to the energy of a proton initially moving with the velocity equal to the average interface velocity,
$\beta_I$, determined from Figure~\ref{fig:beta_int}.
We recognize two distinct acceleration mechanisms below and above the $\beta_I$ threshold, respectively.
Below the threshold, particles ahead of the interface field are accelerated via a prolonged snowplow push.
Since they initially move slower than the interface,
they are trapped in the snowplow field at the front of the interface and accelerated
to their maximum velocity via a relativistic mirror reflection.
This entails that lower initial velocities result in higher final velocities.
Above the threshold, particles outrun the moving accelerating structure.
However, particles starting behind the structure catch up with it and receive an accelerating kick when passing it.
This results in a roughly uniform energy gain for all those initial particles, regardless of their initial energy.
The right panel of Figure~\ref{fig:energy_gain} shows the same plot obtained via the
numerical solution of equations~\eqref{eq:eq_motion_x_snowplow}-\eqref{eq:eq_motion_p_snowplow}.
We notice that substantial energy gain is only achieved in a narrow parameter space
and is characterized by maximizing the acceleration distance,
which is achieved by the particles that are injected in the field as the laser-plasma interface is formed
with an initial energy high enough not to slip all the way through the accelerating field
(roughly $200\,\mev\lesssim \mathcal{E}_I \lesssim 500\,\mev$).
These particle achieve nearly 2.2 GeV energy gain, reaching final energies around 2.5 GeV.

A 1D description of the snowplow acceleration from both our toy-model and Ref.~\onlinecite{liu_accelerating_2022}
shows that the highest energy gain is observed for protons with low initial energies.
These slow protons are effectively reflected, in the reference frame of the moving interface, by the accelerating field, resulting in a significant energy gain.

\begin{figure}[!ht]
\centering
\includegraphics[width=8.6cm]{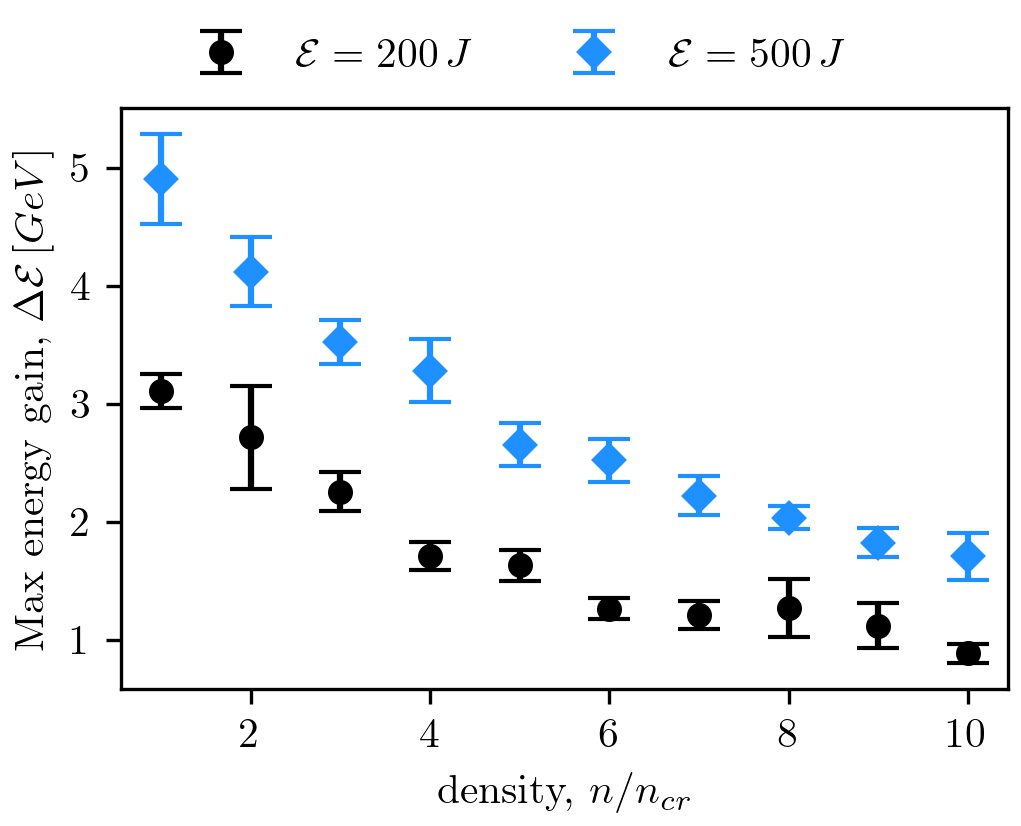}
\caption{The dependence of the maximum proton energy gain on the density of plasma target for
laser energies $200\,\joule$ and $500\,\joule$,
obtained via PIC simulations.}
\label{fig:maxde_vs_density}
\end{figure}

Since we identified the basic features of the proton post acceleration in the snowplow field, i.e.,
the dependence on initial position and energy, we characterize the acceleration process in terms plasma density.
We show the dependence of the maximum proton energy gain on the plasma density in Fig.~\ref{fig:maxde_vs_density}.
As it could have been expected from the waveguide model of high power laser pulse propagation in NCD plasma,
the increase in density leads to the decrease of the maximum proton energy gain, which is connected with
the decrease in the interface velocity and, thus, in acceleration length.
We note that a laser pulse with higher power accelerate protons to higher energies, which is again
due to the interface velocity increasing with laser power. 

\section{Conclusions}\label{sec:conclusions}

In this paper, we explored the possibility of post accelerating a proton beam,
which is generated during the interaction of a multi-PW laser pulses with a NCD target.
Since such an interaction would potentially produce a proton beam with a maximum energy on the order of $1\,\gev$,
but with almost 100\% energy spread, we tested injecting such proton distribution into the second stage,
where a second multi-PW laser beam interacts with a NCD target boosts the energy of these protons.
Our analytical results, based on the waveguide model of high power laser pulse propagation in NCD plasma and a 1D model of snowplow acceleration,
indicate a possibility of a multi-GeV proton energy gain.
The specific energy gain depends on the parameters of laser NCD plasma interaction, in particular on laser power and plasma density. 

We studied the action of the snowplow field on pre-acceleration test protons via
a set of PIC simulations using the code WarpX.
The results of PIC simulations show a multi-GeV energy gain and agree with the analytical estimates.
It was determined that the effectiveness of snowplow post-acceleration depends strongly on
the initial position of test protons and their initial energy.
In other words, the injection scheme should be a crucial component of any future experiment
design on a two-stage proton acceleration.
We found that the maximum energy gain is experienced by those protons that arrive at the second target
before the laser pulse and have the velocity smaller than the laser-plasma interface velocity.
Thus, they are effectively reflected off the moving laser-plasma interface and gain a significant energy boost.
However, when the plasma density increases the effectiveness of the snowplow acceleration decreases leading
to the lower maximum energy gain.
This can be partially mitigated by increasing the laser power,
but due to the weak dependence of laser-plasma interface velocity the overall maximum proton energy gain can not be boosted towards 10's of GeV.
If one envisions a laser-plasma based ion accelerator able to deliver 10's or 100's GeV beams,
then this two stage interaction setup might potentially serve as an injector and booster stages for it.

\section*{Acknowledgments}
This work supported by the U.S. DOE Office of Science Office of HEP under Contract No. DE-AC02-05CH11231,
and used the facilities at the National Energy Research Scientific Computing Center (NERSC) under award HEP-ERCAP0035612.
Lousie Willingale acknowledges support from the National Science Foundation through Award 2408410.
The work of John Palastro and Jessica Shaw is supported by the Office of Fusion Energy Sciences under Award Numbers DE-SC0021057, the Department of Energy National Nuclear Security Administration under Award Number DE-NA0004144, the University of Rochester, and the New York State Energy Research and Development Authority.
The authors would like to acknowledge Marco Garten for useful discussions.

\bibliography{snowplow_biblio}

\end{document}